\documentclass[useAMS,a4paper,usenatbib]{mn2e}

\usepackage{graphicx, array, amssymb,amsmath}
\newcommand{\be}{\begin{equation}}
\newcommand{\ee}{\end{equation}}
\newcommand{\bea}{\begin{eqnarray}}
\newcommand{\eea}{\end{eqnarray}}

\renewcommand\[{\left[}
\renewcommand\]{\right]}

\newcommand\lsim{\mathrel{\rlap{\lower4pt\hbox{\hskip1pt$\sim$}}
    \raise1pt\hbox{$<$}}}
\newcommand\gsim{\mathrel{\rlap{\lower4pt\hbox{\hskip1pt$\sim$}}
    \raise1pt\hbox{$>$}}}
\def\ee{\end{equation}}
\def\be{\begin{equation}}
\newcommand{\Omk}{\Omega_\kappa}

\newcommand{\Omtot}{\Omega_\text{tot}}
\newcommand{\mc}[1]{{\mathcal{#1}}}
\newcommand{\data}{d}
\newcommand{\params}{\theta}
\newcommand{\pone}{\Omk}
\newcommand{\pzero}{\psi}
\newcommand{\mdl}{\mc{M}}

\newcommand{\dr}{\rm{d}}

\newcommand{\Omkfid}{\Omk^*}

\newcommand{\Ode}{\Omega_\text{de}}
\newcommand{\zdec}{z_\text{dec}}
\newcommand{\da}{D_A}
\newcommand{\zbao}{z_\text{bao}}
\newcommand{\weff}{w_\text{eff}} 
\newcommand{\lomk}{o_\kappa}
\newcommand{\erf}{\text{Erf}}

\begin{document}
\setlength{\unitlength}{1mm}

\title[How flat can you get?]{How flat can you get? A model comparison perspective on the curvature of the Universe}

\author[Vardanyan, Trotta and Silk]{Mihran Vardanyan$^{1}$\thanks{E-mail:
mva@astro.ox.ac.uk}, Roberto Trotta$^{1,2}$\thanks{E-mail:
r.trotta@imperial.ac.uk} and Joseph Silk$^{1}$\thanks{E-mail:
silk@astro.ox.ac.uk}\\
$^{1}$Oxford University, Astrophysics,  Denys Wilkinson Building,
Keble Road, Oxford, OX1 3RH, UK \\
$^{2}$Imperial College London, Astrophysics Group, 
	  Blackett Laboratory, Prince Consort Road, London SW7 2AZ, UK}

%\date{Accepted 1988 December 15. Received 1988 December 14; in original form 1988 October 11}

%\pagerange{\pageref{firstpage}--\pageref{lastpage}} \pubyear{2002}

\maketitle

\begin{abstract}

The question of determining the spatial geometry of the Universe is of greater relevance than ever, as precision cosmology promises to verify inflationary predictions about the curvature of the Universe. We revisit the question of what can be learnt about the spatial geometry of the Universe from the perspective of a three--way Bayesian model comparison. By considering two classes of phenomenological priors for the curvature parameter we show that, given current data, the probability that the Universe is spatially infinite lies between 67\% and 98\%, depending on the choice of priors. For the strongest prior choice, we find odds of order 50:1 (200:1) in favour of a flat Universe when compared with a closed (open) model. We also report a robust, prior--independent lower limit to the number of Hubble spheres in the Universe, $N_U \gsim 5$ (at 99\% confidence).  
We forecast the accuracy with which future CMB and BAO observations will be able to constrain curvature, finding that a cosmic variance--limited CMB experiment together with an SKA--like BAO observation will constrain curvature independently of the equation of state of dark energy with a precision of about $\sigma \sim 4.5\times10^{-4}$. We demonstrate that the risk of `model confusion' (i.e., wrongly favouring a flat Universe in the presence of curvature) is much larger than might be assumed from parameter error forecasts for future probes. We argue that a $5\sigma$ detection threshold guarantees a confusion-- and ambiguity--free model selection. Together with inflationary arguments, this implies that the geometry of the Universe is not knowable if the value of the curvature parameter is below $|\Omk| \sim 10^{-4}$. This bound is one order of magnitude larger than what one would naively expect from the size of curvature perturbations, $\sim 10^{-5}$.

\end{abstract}


\section{Introduction}

Constraints on the total energy density of the Universe, $\Omtot$ have improved spectacularly in the last two decades. Before the onset of precision cosmology, the total matter energy content of the Universe was known only with order--of--magnitude precision. The determination of the angular scale of the first acoustic peak in the cosmic microwave background (CMB) was a major milestone towards determining the spatial curvature. The location of the first peak, $\ell \sim 220$, together with estimates of the Hubble constant, implies that the Universe is close to flat. While 10 years ago this statement could be made with an accuracy of order 10\%~\citep{deBernardis:2000gy}, more refined measurements of the CMB power spectrum by WMAP and other experiments have reduced the statistical uncertainty to sub--percent precision in recent years~\citep{Komatsu:2008hk}. In turn, this has allowed us to tighten constraints around a flat Universe with no spatial curvature, $\Omk = 1-\Omtot \sim 0$. This spectacular
increase by over a factor of 100 in accuracy in less than two decades reflects huge steps forward in detector technology, telescope design and computing power. 

As there are only three discrete possibilities for the underlying geometry in a Friedmann--Robertson--Walker Universe\footnote{Although the space of models could be extended to include non--trivial topologies, in this paper we shall keep with the simplest option, namely that the Universe's topology is trivial, as searches for non--trivial topologies have been unsuccessful to date~\citep{Cornish:2003db}.} (namely, flat, close and open, see however~\cite{MersiniHoughton:2007if} for a landscape--motivated alternative with an oscillatory curvature term), the 
question of which one of these three models is the correct description for our Universe is particularly well suited to be phrased in terms of model comparison. In his pioneering application of the Bayesian model comparison framework to cosmology, \cite{Jaffe:1995qu} found that the determination of the Hubble parameter using the Cepheid variables method coupled with a lower limit to the age of Universe already allowed one  to infer that a Universe with vanishing curvature and non--zero cosmological constant was the preferred model (albeit only with modest odds of about 7:1). More recently, the WMAP 3--year data implied that the odds in favour of a flat Universe increased to between 29:1~\citep{Trotta:2005ar} and 48:1~\citep{Kunz:2006mc} when comparing a flat Universe with curved models (both open and closed).

Unlike many dark energy models that are mostly phenomenological, models predicting curvature are rooted in fairly well understood physics, a feature which helps in setting physically motivated priors on the model parameters. For example, the possibility of a flat, $\Omk \sim 0$ Universe has long been
favoured by theoretical prejudice, as a flat or close--to--flat Universe is a generic prediction of the inflationary scenario, which appears to have been confirmed by observations to date. With the prospect of even more vigorous observational campaigns
coming up in the next decade, it is timely to ask to which
point the accuracy in $\Omk$ can and should be pushed before the
question about the flatness of the Universe becomes irrelevant,
uninteresting or undecidable. Determining curvature is also important in order to avoid mistaking a non--flat Universe for an indication of an evolving dark energy density, see e.g.~\cite{Knox:2006ux,Clarkson:2007bc,Virey:2008nu}.

In this paper we address the capability of future CMB and baryonic acoustic oscillations (BAO) observations to constraint curvature, both from the point of view of parameter constraints and from the perspective of a three--way Bayesian model comparison. We are primarily interested in the accuracy that can be achieved using the acoustic scale as a standard ruler, although complementary observations (for example, supernova type Ia (SNIa) or weak lensing observations) will help to break existing degeneracies between curvature and the dark energy equation of state~\citep{Clarkson:2007bc}, thereby improving the statistical power. A fundamental limit to our ability to determine curvature is set by the order of magnitude of local fluctuations in the spatial curvature, $\Delta \Omtot \sim 10^{-5}$. We investigate how this translates into terms of model selection and, crucially, model confusion, and show that the size of the fluctuations means that the question of curvature becomes statistically undecidable for $|\Omk| \lsim 10^{-4}$, i.e.~about one order of magnitude above the naive expectation, and this is regardless of the amount of data gathered.

This paper is organized as follows. In section~\ref{sec:setup} we introduce the data we use and our forecast procedure, while we briefly review relevant aspects of Bayesian model selection in section~\ref{sec:modsel}, where our prior choices are discussed. We present the evidence from current data in section \ref{sec:today}, while section \ref{sec:tomorrow} gives the results of our forecast for future probes and discusses model confusion. We give our conclusions in section~\ref{sec:conclusions}.

\section{Setup and methodology}
\label{sec:setup}

\subsection{Measuring the acoustic scale}

The acoustic peaks in the CMB power spectrum measure the projected sound horizon at recombination. The comoving sound horizon at decoupling is given by 
\be
r_s(\zdec) = \frac{c}{H_0} \int_{\zdec}^\infty \frac{c_s}{H(z)} \dr z ,
\ee
where $H_0$ is the Hubble constant today, $\zdec$ is the redshift of decoupling and $c_s$ is the sound speed of the coupled photon--baryon fluid, 
\be
c_s = \frac{1}{\sqrt{3(1+R)}}, 
\ee 
with $R = 3\rho_b/\rho_\gamma \approx [670/(1+z)](\Omega_b h^2/0.022)$. Here, $\rho_b$ and $\rho_\gamma$ are the time--dependent energy densities of baryons and photons, respectively, while $\Omega_b$ is the energy density parameter for baryons today. The function $H(z)$ is given by
\begin{multline}
%\nonumber
H^2(z)= \Big( \Omega_m(1+z)^3+\Omega_r(1+z)^4+\Omk (1+z)^2\\
+\Ode \exp \left ( 3\int_{0}^{z}\frac{1+w(x)}{1+x} \dr x \right ) \Big),
\end{multline}
where the dark energy time evolution is described by the present--day dark energy density in units of the critical density, $\Ode$, and by its equation of state, $w(z)$. The energy density parameter for radiation (photons and neutrinos, taken here to be massless) is 
$\Omega_r = \frac{\pi^2}{15}[ 1+(21/8)(4/11)^{4/3}]T_\text{CMB}^4/h^2 \approx 4.13\times10^{-5}/h^2$, while 
\be \label{eq:curvp}
\Omk = -\frac{\kappa c^2}{a_0^2 H_0^2}
\ee
is the curvature parameter ($a_0$ is the scale factor today). The curvature constant $\kappa$ determines the geometry of spatial sections: $\kappa = 0$ for a flat Universe, $\kappa = +1$ for a closed Universe and $\kappa = -1$ for an open Universe. 

The comoving distance to an object at redshift $z$ is given by 
\be \label{eq:comovingdistance}
\chi(z) = \frac{c}{H_0 a_0}  \int_0^z \frac{\dr x}{H(x)}.
\ee 
Given knowledge of the comoving length $\lambda$ of an object at redshift $z$, measurement of the angle subtended by it on the sky, $\theta$, determines its angular diameter distance, $\da(z)$
\be
\da(z) = \frac{\lambda (a/a_0)}{\theta} = \frac{a_0 S_\kappa(\chi)}{1+z},
\ee
where $S_\kappa(y)$ is $y$, $\sin(y)$ or $\sinh(y)$ for $\kappa=0,+1,-1$, respectively. A number of authors~\citep{Bond:1997wr,Bowen:2001in,Melchiorri:2000px,Kosowsky:2002zt,Jimenez:2004ct} have pointed out that the morphology of the acoustic peaks in the CMB power spectrum is largely controlled by the baryon density $\Omega_b h^2$ and by two `shift parameters'
\begin{align}
l_a & \equiv \pi \chi(\zdec)/r_s(\zdec), \\	
R &\equiv \sqrt{\Omega_m H_0^2}\chi(\zdec)/c.
\end{align}
In the context of the recent interest in dark energy, the usefulness of employing both shift parameters (and their correlations) as a handy summary of CMB constraints has been brought into sharp focus by~\cite{Wang:2007mza}. In this work, we follow their method of employing `distance priors'  as constraints on $(l_a, R, \zdec)$ for a summary of the information given by the CMB on the expansion history of the Universe.

At lower redshift, the acoustic signature has been recently detected in the distribution of galaxies \citep{Percival:2007yw,Eisenstein:2005su,Gaztanaga:2008xz}, thereby providing further constraints on the recent expansion history of the Universe. Future large galaxy surveys are expected to considerably improve present--day accuracy, by simultaneously determining the angular diameter distance and the Hubble function $H(z)$, which can be obtained by measuring the acoustic scale in the radial direction if spectroscopic data are available. This is because the radial extent of a feature along the line of sight is related to the redshift range $\Delta z$ by  
\begin{equation}
\label{parallel}
r_\parallel = \frac{c\Delta z}{H(z)},
\end{equation}
hence a measurement of $r_\parallel $ allows a direct reconstruction of $H(z)$.

\subsection{Parameters and data sets}

In this paper, we consider cosmologies containing baryons, cold dark matter, dark energy and a possible curvature term. The radiation density is fixed to the appropriate value for 3 families of massless neutrinos throughout. Dark energy is taken to be either in the form of a cosmological constant, $w=-1$, or is described in terms of an effective equation of state $\weff \neq -1$, which is taken to be constant with redshift. Of course more complex parameterizations are possible, and in particular one could consider an evolving dark energy equation of state which changes with redshift (for constraints on such models, see e.g.~\cite{Zunckel:2007jm}).  A particularly popular phenomenological parameterization of a time--evolving dark energy is to describe the equation of state as $w(z) = w_0 + \frac{z}{1+z}w_a$, with two free parameters $(w_0, w_a)$. We comment below on the impact that adopting such a more general dark energy model would have on our results.

We employ a Metropolis--Hastings Markov Chain Monte Carlo procedure to derive the posterior distribution for the parameters in our model. We take flat priors on the following quantities: $\Omega_m h^2, \Omega_b h^2, l_a, R, \weff$ (whenever the latter is not fixed to $-1$). The prior bounds on the first 3 parameters are irrelevant, as the posterior is well constrained within the prior. For $\weff$ we take a prior range $-2 \leq \weff \leq -1/3$, with the lower bound cutting off some of the posterior for some of our data combinations (see below). Finally, the choice of prior for $\Omk$ is fundamental for the model comparison part, and we discuss it in detail in section~\ref{sec:priorchoice}.

When considering present--day data, we include the WMAP 5--year data~\citep{Dunkley:2008ie} via their constraints on the shift parameters and the baryon density, following the method employed in~\cite{Komatsu:2008hk}. We also include the SDSS baryonic acoustic scale measurement as an additional datum at redshift $z=0.35$ by adding a Gaussian distributed measurement of the quantity
\be
 A = \left( \chi^2(\zbao) \frac{c \zbao}{H(\zbao)}\right)^{1/3}\frac{\sqrt{\Omega_m H_0^2}}{c\zbao}, 
\ee
where $\zbao = 0.35$. We employ the mean value $A=0.474$ with standard deviation $\sigma_A = 0.017$~\citep{Eisenstein:2005su}. We also add the Hubble Key Project determination of the Hubble constant today, as a Gaussian datum with mean $H_0 = 72$ km/s/Mpc and standard deviation $\sigma_{H_0} = 8$ km/s/Mpc~\citep{Freedman:2000cf}. SN type Ia data are included in the form of the Supernovae Legacy Survey (SNLS) sample~\citep{Astier:2005qq}. 

\subsection{Future data} 

We now turn to describe our procedure for simulating constraints from future CMB and BAO observations. 

\subsubsection*{CMB data: Planck and CVL experiment}

We consider two types of future CMB measurements, one from the Planck satellite (due for launch early in 2009), which will measure the temperature power spectrum with cosmic variance accuracy up to $\ell \sim 2000$, and will considerably improve current precision in the ET and EE power spectra. We also consider a hypothetical Cosmic Variance Limited (CVL) experiment, which would measure the TT, EE and ET spectra with cosmic variance precision up to $\ell = 2000$. This is meant to represent the ultimate precision obtainable from measurements of the acoustic scale at recombination (although of course extra information on the expansion history will be available, e.g. via the integrated Sachs--Wolfe effect or CMB lensing. As mentioned above, we are concerned with the accuracy achievable by `geometric' means alone). 

We start by choosing a fiducial value of the cosmological parameters around which to generate simulated CMB data. We employ $\Omega_b h^2 = 0.02268, \Omega_\text{cdm} h^2 = 0.1081, \Omk=0, \weff = -1$, which are in good agreement with the current best--fit from WMAP and other CMB observations (the values of the spectral tilt and perturbations normalization are irrelevant for our analysis as we only employ effective distance measures to the last scattering surface from the CMB). The corresponding CMB power spectra are computed using the CAMB code~\citep{Lewis:1999bs}. We then add noise according to the procedure described in~\cite{Lewis:2005tp}, with noise levels appropriate for either Planck or the CVL experiment (which has no noise up to $\ell=2000$). Finally, a modified version of \texttt{CosmoMC}~\citep{Lewis:2002ah} is employed to fit the resulting noisy power spectra and to recover the covariance matrix for the parameters $(R, l_a, \zdec)$, following the method described in~\cite{Komatsu:2008hk}, which shows that constraints on this set of parameters are essentially equivalent to constraints on  $(R, l_a, \Omega_b h^2)$. \cite{Li:2008} have analyzed in detail the loss of information involved in going from the full CMB data analysis to the use of the constraints on the set $(R, l_a, \zdec)$ and have found that the covariance matrix method represents accurately the information contained in the CMB. \cite{Mukherjee:2008kd} investigated the application of this formalism to Planck priors, and found a significant correlation between the shift parameters and the spectral tilt, $n_S$. In this work, we do not include the tilt in the description of Planck data, on the basis that we never use Planck data alone to derive our constraints on the curvature parameter. Thus, the degeneracy between $n_S$ and the shift parameters can be assumed to be effectively broken when including non--CMB observations, in particular data on the matter power spectrum which, by extending very considerably the lever arm of the CMB, are expected to be able to reduce the uncertainty on $n_S$ to a level which does no longer impact on the accuracy of the shift parameters. 

 The fiducial values for our reference choice of parameters are $(R, l_a, \zdec) = (302.06, 1.709, 1090.46)$. The corresponding covariance matrices for Planck and the CVL experiment are given in Table~\ref{table:covmats}. 

\begin{table}
\begin{center}\begin{tabular}{l | ccc}
\hline&  \multicolumn{3}{c}{Planck}  \\
  	&$ l_a $& $R$ & $\zdec$   \\\hline
$ l_a $& $5.96\cdot10^{-3}$& $  1.96\cdot10^{-4}$& $  1.02\cdot10^{-2}$ \\
$ R $& & $  2.15\cdot10^{-5}$& $  1.06\cdot10^{-3}$ \\
$ \zdec $& & & $  6.90\cdot10^{-2}$
\\\hline	
\hline&  \multicolumn{3}{c}{Cosmic Variance Limited (CVL)}  \\
  	&$ l_a $& $R$ & $\zdec$   \\\hline
	 $ l_a $& $8.12\cdot10^{-4}$& $  3.89\cdot10^{-5}$& $  1.35\cdot10^{-3}$ \\
 $ R $& & $  6.23\cdot10^{-6}$& $  2.52\cdot10^{-4}$ \\
 $ \zdec $& & & $  1.47\cdot10^{-2}$ \\\hline
			  
\end{tabular} \caption{Covariance matrices for the distance parameters  $(l_a, R, \zdec)$ for Planck (top) and the CVL experiment (bottom). \label{table:covmats}}
\end{center}
\end{table}

In obtaining the covariance matrix for Planck and the CVL experiment, the curvature parameter has been allowed to vary (with a flat prior over a suitably large range so that the posterior is much narrower than the prior), in order to obtain errors that correctly account for degeneracies in $\Omk$. On the other hand, the equation of state parameter has been fixed at $w=-1$ when computing the covariance matrix. This is expected to be irrelevant as the whole point of using CMB `distance priors' of this sort is precisely that they are largely independent on the assumed dark energy model (at least as long as the contribution of dark energy in the early Universe is negligible, which is the case here since we never fit evolving dark energy models).

This covariance matrix is then used as the CMB high--redshift constraint. Notice that although the simulated data are obtained around a flat fiducial model,  we can safely use the resulting covariance matrix to represent CMB distance priors even when the fiducial model is slightly changed to $\Omk \neq 0$ (as long as the change is not too large as to radically modify degeneracy directions), as we do below, where we employ fiducial models with $|\Omk|\leq5\times 10^{-3}$.

\subsubsection*{BAO data: WFMOS and SKA--like experiment}

Regarding future BAO measurements, we adopt two benchmark experiments. One is the Wide--Field Multi--Object Spectrograph (WFMOS), a proposed instrument for the 8--m Subaru telescope which will employ a fiber--fed spectrograph to carry out a low ($z\sim1$) and a deep ($z\sim3$) survey to determine the acoustic oscillation scale both in the transverse and in the radial direction~\citep{Bassett:2005kn}. WFMOS could be operating around 2015. We also consider a more futuristic type of measurement, of the kind that could be delivered by the Square Kilometer Array (SKA) radiotelescope around 2020 by performing a full--sky survey of HI emission. 

In modeling the
accuracy of these observations, we closely follow the treatment of~\cite{Blake:2005jd}, to which we refer for full details. In summary, we employ the following fitting formula for the fractional accuracy of the determination of the transverse and radial acoustic scale:
\begin{equation}
\label{eq:fittingformula}
x =   x_0 \sqrt{\frac{V_0}{V}}\left(1 + \frac{n_\text{eff}}{n} \frac{D(z_0)^2}{b_0^2D(z)^2}\right)f(x)
\end{equation}
with  $f(z) = (z_m/z)^\gamma$ for $z<z_m$ and $f(z) = 1$ otherwise. Here, $x$ is the fractional accuracy in the determination of either $\chi(z)/r_s(\zdec)$ (transversal direction) or $cH(z)^{-1}/r_s(\zdec)$ (radial direction) which can be obtained by a spectroscopic survey of volume $V$, measuring a galaxy density $n$ at redshift $z$. In the above equation, $D(z)$ is the growth factor, $(V_0,  z_0, b_0)$ are the values for a reference survey while $(x_0, n_\text{eff}, z_m, \gamma)$ are fitted parameters obtained via a simulation study by~\cite{Blake:2005jd}, which depend on whether one is considering a measurement of the acoustic scale in the radial or tangential direction. We employ the values given in Table 1 of~\cite{Blake:2005jd} for a spectroscopic survey, as appropriate for WFMOS and the SKA. In Eq.~\eqref{eq:fittingformula} we recognise a term $\propto 1/\sqrt{V}$ representing the scaling of the number of available Fourier modes with volume, a term $\propto 1/(nD^2)$ representing shot noise and a redshift--dependent cut--off term $\propto 1/z^\gamma$ below $z_m = 1.4$ that suppresses non--linear modes (which however might also be included in a full non--linear analysis, thereby considerably increasing the BAO constraining power, see e.g.~\cite{Crocce:2007dt}).  

The WFMOS parameters are taken from the results of the detailed optimization study by~\cite{Parkinson:2007cv}. Although~\cite{Parkinson:2007cv} optimized WMFOS experimental parameters for dark energy constraints in a flat Universe, we expect that their general preference for a low redshift bin with as large as possible an area would still hold true even in an optimization scenario where curvature is allowed to vary. For definiteness, we adopt the values given in Table 2, column B of~\cite{Parkinson:2007cv}. This gives a wide bin at low redshift, covering an area of $A_\text{low} = 5600$ deg$^2$ at a median redshift $z_\text{low} = 1.08$, a redshift width $\Delta z_\text{low} = 0.35$ and a number density of galaxies $n_\text{low} = 7.1 \times 10^4$ ($h^3$/Mpc$^3$). The high--redshift bin has parameters  $A_\text{high} = 150$ deg$^2$, $z_\text{high} = 3.15$, $\Delta z_\text{high} = 0.13$ and $n_\text{low} = 0.13 \times 10^4$ ($h^3$/Mpc$^3$). We have found that essentially all of the constraining power of this configuration comes from the $z\sim1$ bin, in agreement with the results of other studies.  

The SKA is still in the design phase, hence its precise performance is somewhat uncertain at the moment (see e.g.~\cite{Blake:2004pb} for an overview). We choose to represent its capabilities by assuming measurements of both the transverse and radial acoustic scales equally spaced in 4 redshift bins at $z=1,2,3,4$, each of width $\Delta z = 0.4$. We further assume that the SKA will survey the whole sky ($A = 20,000$ deg$^2$) and that the density of galaxies will be large enough as to be able to neglect the shot noise term (i.e., $nP > 3$, where $P$ is the power of the fluctuations). Some of these choices are somewhat optimistic and further detailed modeling is required in order to be able to verify the capability of the SKA to achieve these specifications. However, we have taken here the SKA to represent a sort of `ultimate' BAO measurement, which provides with a flavour of what the ultimate level of accuracy of the method might be. We add everywhere an external Gaussian constraint on the value of the Hubble parameter corresponding to the HST Key Project determination,  $H_0 = 72\pm8$ km/s/Mpc~\citep{Freedman:2000cf}.

Of course we are only dealing with statistical uncertainties here, and the issue of systematics will at some point have to be addressed in detail, as the statistical error becomes smaller. However, BAO are particularly promising in this respect, thanks to the very low level of systematics expected (e.g., \cite{Trotta:2006gx}).

\section{Curvature and Bayesian model comparison}
\label{sec:modsel} 

Determining whether the Universe is flat or not is one of the most interesting questions in modern cosmology. This is however not a problem of parameter constraints, but rather of model comparison. In this section, we briefly describe Bayesian model comparison and its use in forecasting the power of future observations (for more details, see e.g.~\cite{Trotta:Bayes,Trotta:BMIC}). We then discuss the choice of priors on $\Omk$ and motivate it in the light of theoretical considerations. 

\subsection{Model comparison}

From Bayes' theorem, the probability of model $\mc{M}$ given
the data, $p(\mc{M}|d)$, is related to the Bayesian evidence (or model
likelihood) $p(d|\mc{M})$ by
\begin{eqnarray} \label{eq:modelpost}
 p(\mc{M}|d)&=&\frac{p(d|\mc{M})p(\mc{M})}{p(d)}\, ,
\end{eqnarray}
where $p(\mc{M})$ is the prior belief in model $\mc{M}$, $p(d)=\sum_i
p(d|\mathcal{M}_i)p(\mc{M}_i)$ is a normalization constant and
\be \label{eq:Bayesian_evidence}
p(d|\mathcal{M}_i)=\int\!d\theta\, p(d|\theta, \mc{M}_i) p(\mc{M}_i)
\ee
is the Bayesian evidence. Given two competing models $\mc{M}_1,
\mc{M}_2$, the Bayes factor $B_{01}$ is the ratio of the models'
evidence
\begin{eqnarray}
 B_{01}&\equiv&\frac{p(d|\mc{M}_0)}{p(d|\mc{M}_1)}\, ,
\end{eqnarray}
where large values of $B_{01}$ denote a preference for $\mc{M}_0$,
whereas small values of $B_{01}$ denote a preference for
$\mc{M}_1$. The `Jeffreys' scale' (Table~\ref{tab:jeff}) gives an empirical scale for
translating the values of $\ln B_{01}$ into strengths of belief (following the prescription given in~\cite{Gordon:2007xm} for denoting the different levels of evidence). Recently, the framework of model comparison has been extended to include the possibility of `unknown models' discovery~\citep{doubt}.

\begin{table}
 \begin{tabular}{l l  l} 
  $|\ln B_{01}|$ & Odds & Strength of evidence \\\hline
 $<1.0$ & $\lsim 3:1$  & Inconclusive \\
 $1.0$ & $\sim 3:1$ & Weak evidence \\
 $2.5$ & $\sim 12:1$  & Moderate evidence \\
 $5.0$ & $\sim 150:1$ & Strong evidence \\
 \hline
\end{tabular}
\caption{Empirical scale for evaluating the strength of evidence when
comparing two models, $\mdl_0$ versus $\mdl_1$ (so--called
`Jeffreys' scale'). The right--most column gives our
convention for denoting the different levels of evidence above
these thresholds.\label{tab:jeff} }
\end{table}

Given two or more models, it is straightforward (although often
computationally challenging) to compute the Bayes factor. Several numerical algorithms are available today to compute the Bayesian evidence. Recently, a very effective algorithm, called `nested sampling' \citep{SkillingNS,Skilling:2006}, has become available, which has been
implemented in the cosmological context
by~\cite{Bassett:2004wz,Mukherjee:2005wg,Shaw:2007jj,Feroz:2007kg,Bridges:2006mt}. Here we are interested in the simpler scenario where the two models are {\em nested}, i.e. where the more complicated model reduces to the simpler one for a specific choice of the extra parameter. In our case, the extra parameter is the curvature, $\Omk$, with a curved Universe reverting to a flat one for $\Omk=0$.
Writing for the extended model parameters $\params =
(\pzero, \pone)$, where the simpler (flat) model $\mdl_0$ is obtained by
setting $\pone = 0$, and assuming further that the prior is
separable (which is the case here), i.e.\ that
 \begin{equation}
 p(\pzero, \pone| \mdl_1) = p(\pone | \mdl_1) p(\pzero | \mdl_0),
 \end{equation}
the Bayes factor can be written in all generality as
 \begin{equation} \label{eq:savagedickey}
 B_{01} = \left.\frac{p(\pone \vert \data, \mdl_1)}{p(\pone |
 \mdl_1)}\right|_{\pone = 0}.
 \end{equation}
This expression is known as the Savage--Dickey density ratio
(SDDR, see \cite{Verdinelli:1995} and references therein. For cosmological applications, see~\cite{Trotta:2005ar}). The
numerator is simply the marginal posterior for $\Omk$, evaluated at the flat Universe value, $\Omk = 0$, while the
denominator is the prior density for the model with $\Omk\neq0$,
evaluated at the same point. This technique is particularly useful
when testing for one extra parameter at a time, because then the
marginal posterior $p(\pone \vert \data, \mdl_1)$ is a
1--dimensional function and normalizing it to unity probability
content only requires a 1--dimensional integral, which is computationally simple to do.

\subsection{A three--way model comparison}
\label{sec:priorchoice}

We consider each possible choice of the curvature parameter $\kappa$ as defining a separate model. This means that we perform a three--way model comparison between a flat ($\kappa=0$), an open ($\kappa=-1$) and a closed ($\kappa=+1$) Universe. It is obvious that we might want to distinguish between a flat Universe and non--flat alternatives. However, it is also convenient to separate the positive and negative curvature scenarios as two different models. This will allow us to make statements on the probability that the Universe is finite (corresponding to the closed case), and also to consider in a natural way a prior on $\Omk$ that is flat in the log of the curvature parameter, as motivated below. 

For the prior probability assigned to each of the three possible geometries, we make a non--committal choice of assigning equal probabilities to each, i.e. $p(\mdl_i) = 1/3$ ($i=-1,0,+1$), where the labels of the models gives in each case the value of $\kappa$. Of course, different choices are possible: for example, if one feels that inflation strongly motivates an almost flat Universe, this might be reflected by increasing the value of $p(\mdl_0)$ (we comment further on this below). It is straightforward to include such a theoretical preference by recalibrating our results if one wanted to.

From the definition of the model's posterior probability, Eq.~\eqref{eq:modelpost} and as a consequence of our assumption of equal prior probabilities for our models, we obtain for the posterior probability of the flat model the handy expression
 \be \label{eq:M0post}
 p(\mdl_0 |d) = \frac{1}{1+B_{01}^{-1} + B_{0-1}^{-1}}. 
\ee
The posterior probabilities of the $\kappa\neq0$ models can easily be obtained by suitably exchanging the indexes of the Bayes factors. 

Each one of the models is described by a 6 parameter vanilla $\Lambda$CDM model (or a 7--parameter dark energy model with $\weff\neq -1$). In principle we need to specify the priors on these parameters, too, but since they are common parameters to all models, their priors effectively cancel, as shown above by Eq.~\eqref{eq:savagedickey}. Whenever we include the extra parameter $\weff \neq -1$, we always add it to all models at the same time, therefore the model comparison is always only about the curvature.

Model selection relies on a choice of prior for the extra parameter in the more complex model, which controls the strength of the Occam's razor effect, in our case $\Omk$. Such a choice should be motivated by physical considerations, ideally stemming from the theoretical properties of the model under scrutiny (see~\cite{Efstathiou:2008ed} for a critical view). We therefore need to consider carefully the prior distribution for the value of the parameter describing the curvature of spatial sections for the non--flat models. 

\subsection{Priors on the curvature parameter}

 A possible parameterization of the spatial curvature is given by the curvature parameter today, Eq.~\eqref{eq:curvp}. 
A flat Universe ($\Omk = 0$) would therefore appear to be a point null hypothesis, to be tested against a more complex alternative model (with $\Omk \neq 0$). In the context of inflation, however, the geometry needs not be exactly flat. Indeed, the whole point of inflation is precisely to provide a mechanism to avoid such an implausible fine tuning. For $\kappa \neq 0$, inflation ensures that the curvature scale tends to zero:
 \be \label{eq:efolds}
 | \Omk | \approx \exp(-2N_b)
 \ee
where $N_b$ is the number of e--folds before our current comoving Hubble volume exited the horizon (see e.g.~\cite{Tegmark:2004qd}). If we had a measure for the parameter space of inflationary potentials (for example from string theory), we could in principle convert the probability distribution for the potential into a prior on $N_b$, and from here into a prior on $\Omk$. This is not necessary in practice, because local fluctuations in our Hubble volume limit the precision to which we can observe deviations from $\Omk=0$ to $\sim 10^{-5}$ (see \cite{Waterhouse:2008vb} for a more rigorous motivation of this result). Therefore, provided $N_b \gsim 5.8$, inflationary predictions are observationally indistinguishable from a flat Universe. Given that $N_b$ could be anything between 0 and $\infty$, it appears to be a reasonable approximation to neglect models with $N_b < 5.8$ (see~\cite{Tegmark:2004qd} for a justification), although such cases could be considered as a particular class of models if one wanted. For definiteness, in the following we will take the inflationary prediction to correspond to $|\Omk| < 10^{-5}$, thereby extending the point null hypothesis that $\kappa = 0$ to include such small values of the curvature parameter. Because of the fundamental limitation of cosmic variance, we argue that it is pointless to consider the prior distribution of $\Omk$ below the threshold value of $10^{-5}$. 

In summary, we describe a generic inflationary prediction as being $|\Omk| < 10^{-5}$ (with no free parameters) and a prior model probability $p(\mdl_0) = 1/3$. The latter assignment could of course be amended if one felt that inflation is compellingly motivated by its ability to solve other problems such as the homogeneity and monopole problems, in which case the prior probabilities for non--inflationary models would have to be correspondingly reduced\footnote{An important point is that we are here neglecting the possibility of inflationary models predicting, for example, closed Universes with sizeable values of the curvature parameter (see e.g.~\cite{Lasenby:2003ur} for such a model). So what we describe as a generic inflationary prediction is really only a subclass of possible inflationary scenarios. It would be simple to extend the model comparison to include other subclasses of inflationary models if one wanted to.}.  However, this is not essential for what follows, as we will mostly quote Bayes factors which give the {\em change in degree of relative belief} between two models in the light of the data. This means that the model's prior specification has no influence on the Bayes factor. In any case, it is straightforward to propagate a change in the models' prior probability to the model posterior probabilities that we give below.

The model comparison is then fully defined once we choose a prior pdf for the extra parameter in the curved models, for values $|\Omk| > 10^{-5}$. The prior should reflect our state of belief on the possible values of the relevant parameter before we see the data. We adopt two different priors choices for deviations from flatness, representing two different states of beliefs about the locus of possibilities for the geometry of the Universe.
 \begin{itemize}
 \item {\bf Flat prior on $\Omk$: the `Astronomer's prior'.} This is prior is motivated by considerations of consistency with mildly informative observations on the properties of the Universe. Requiring that the Universe is not empty gives $\Omk > -1$, barring the exotic case of a negative cosmological constant. The age of a Universe containing only matter can be approximated by $t_0 H_0= (1 + \Omtot^{0.6}/2)^{-1}$, which means that a positively curved Universe is increasingly at odds with the age of the oldest objects, requiring $t_0 \gsim 10$ Gyr. A positive cosmological constant helps to solve the age problem, but if $\Omtot \gsim 2$ then $t_0 \lsim 8 h$ Gyr even in a de Sitter Universe.  So unless $h\gg1$ a Universe with $\Omk\gsim 1$ is too young even in the presence of $\Lambda$. The lower limit for the curvature parameter is given by $|\Omk| = 10^{-5}$ as discussed above. However on a linear scale this is effectively equivalent to setting the lower limit to 0. These considerations therefore lead to the prior choice:  
\begin{align}
p_A(\Omk | \mdl_1) = 1 & \quad {\rm for } & -1  \leq \Omk \leq 0 \\
p_A(\Omk | \mdl_{-1}) = 1 & \quad {\rm for }  & 0 \leq \Omk \leq 1, 
\end{align} 
where the subscript $A$ denotes that this prior is based on the astronomical considerations sketched above. 
 \item {\bf Flat prior on $\ln\Omk$: the `Curvature scale prior'.}
 Alternatively, we might consider the curvature scale today: 
 \be 
 a_0   = \frac{c}{H_0}\[ \frac{\kappa}{\Omtot -1}\]^{1/2} 
 \ee 
Clearly, a flat prior on $\Omk$ does not correspond to a flat prior on $a_0$, as the two pdf's are related by 
 \be
 p(a_0) = p(\Omk)   {\Big\arrowvert}\frac{\dr \Omk}{\dr a_0}{\Big\arrowvert}.
 \ee
Hence a flat prior on $\Omk$ gives an informative prior on the curvature scale, $p(a_0) \propto a_0^{-3}$, which prefers more strongly curved Universes. A flat prior on $\ln\Omk$ represents a state of belief which is indifferent with respect to the order of magnitude of the curvature parameter. It is easy to see that this implies a similar state of indifference on the order of magnitude of the curvature scale, since a flat prior on $\ln\Omk$ is flat on $\ln a_0$, as well. Furthermore, such a prior is also flat in the number of e--folds, as a consequence of Eq.~\eqref{eq:efolds}. The upper cutoff for the prior can be established by requiring that the curvature scale be larger than the Hubble horizon radius, $H_0^{-1}$. Furthermore, we are free to choose the basis in which the logarithm is taken, and in the following we shall employ base 10 logarithms. We thus define the variable
\be
 \lomk \equiv \log_{10}|\Omk|.
 \ee
These considerations lead to the prior choice 
 \be
p_C(\lomk | \mdl_\kappa) = 1/5  \quad {\rm for }  -5 \leq \lomk \leq 0
\ee 
for $\kappa=-1,1$ and where the subscript $C$ denotes that this prior is based on a state of indifference with respect to the curvature scale. 
 \end{itemize}
When employing the SDDR to evaluate the Bayes factor between a flat and a curved model for the curvature scale prior, we evaluate the marginal posterior and the prior of Eq.~\eqref{eq:savagedickey} at the value $\lomk = -5$ (corresponding to $|\Omk| = 10^{-5}$), since this is the value at which the curved models revert to a flat Universe for our choice of priors.  

Other choices of parameterization for curvature (and the associated priors) are certainly possible and might be well motivated from a theoretical point of view. For example,~\cite{Adler:2005mn} introduce a constant flatness parameter $\epsilon$ given by the ratio of two fundamental constants determining the dynamics of the expansion, and show that the value of $\epsilon$ is in many ways a better indicator of ``fine tuning'' than $|\Omk|$. Again, if one had access to the distributional properties of the fundamental constants and from there to the distribution of $\epsilon$, one could imagine building a physically motivated prior on that quantity instead. However, since presently we are unable to predict from first principles the distributional properties of such quantities, we prefer to adopt a semi--phenomenological approach, informed by the physical reasoning sketched above.

\subsection{Implications for the number of Hubble spheres and the size of the Universe} 

For closed Universes (i.e., for $\mdl_1$), it is interesting to translate the probability distribution for $\Omk$ or $\lomk$ into the corresponding posterior for the number of particle horizon volumes that fit into the current spatial slice. Following~\cite{Scott:2006kga}, we thus define 
\be
N_U \equiv \frac{2 \pi}{2 \chi - \sin(2\chi)}
\ee
as the ratio of the present volume of the spatial slice to the apparent particle horizon (assuming radiation domination into the infinite past), where $\chi$ is the comoving radial distance defined in Eq.~\eqref{eq:comovingdistance}, and for closed models $0\leq \chi \leq \pi$. Given our choice of priors, we can easily translate the results of the previous section into the posterior for $N_U$. Clearly, under either the flat or open models ($\mdl_0$ or $\mdl_{-1}$), the volume of the spatial slice is infinite and hence $N_U$ goes to infinity. In the Bayesian framework, we can give the probability that this is the case, namely that we live in an infinite Universe. For our choice of model priors, it follows that
\begin{align} \label{eq:post_prob_infinity}
p(N_U = \infty | d) & = p(\mdl_0|d) + p(\mdl_{-1}|d)  \\
& = p(\mdl_0|d) \left(1 + \frac{1}{B_{0-1}} \right),
\end{align}
where $p(\mdl_0|d)$ is given by Eq.~\eqref{eq:M0post}. For other choices of model priors (for example, $p(\mdl_0) \gg 1/3$, representing a stronger degree of theoretical prejudice in favour of inflation) one should rescale the posteriors accordingly.

\subsection{Model comparison forecasting} 

When considering the capability of future probes it is customary to quantify their expected performance in terms of a `Figure of merit' (FOM). Several FOMs exist, but they mostly focus on the parameter constraint capabilities of  future observations (e.g., ~\cite{Bassett:2004st}). However, many (and indeed perhaps most) questions of interest are actually about model comparison: for example, determining whether dark energy is a cosmological constant, or whether the Universe is flat are clearly model comparison problems. FOMs geared for parameter constraints capabilities do not necessarily reflect the model comparison potential of a future probe (see~\cite{Liddle:2007ez} for details).  

A few techniques have been put forward to assess the model comparison capability of future observations:~\cite{Trotta:2007hy} has introduced a technique called PPOD, which computes the probability distribution of the outcome of a future model comparison; \cite{Pahud:2006kv,Pahud:2007gi} have looked at the ability of Planck to obtain a decisive model selection result regarding the spectral index; \cite{Liddle:2006kn} have applied a similar technique to the problem of distinguishing between an evolving dark energy and a cosmological constant.

Here we adopt a procedure similar in spirit to~\cite{Liddle:2006kn}. We want to quantify the ability of future CMB and BAO measurements to obtain a correct model selection outcome about the geometry of the Universe. We therefore simulate data as explained above for three different fiducial values of $\Omk$: for a flat model, $\Omkfid = 0$, and for two different closed models, $\Omkfid = -10^{-3}$ and $\Omkfid = -5 \times 10^{-3}$. From the posterior distribution obtained from simulated data, one can compute the corresponding Bayes factor via the SDDR,~Eq.~\eqref{eq:savagedickey}. Once interpreted against the Jeffreys' scale, the future Bayes factor then allows us to determine whether the experiment will be accurate enough to correctly identify the true model, and if so with what strength of evidence. Our procedure is thus similar to the one adopted in~\cite{Pahud:2007gi,Pahud:2006kv}.
 
In principle one could repeat the forecast for several other values of $\Omkfid$, thus more densely covering  the range of possible fiducial values. However, we found that these 3 cases are representative of 3 interesting possibilities. The case $\Omk = -5 \times 10^{-3}$ has been chosen because it lies just below current combined limits from CMB, BAO and SNIa, and within reach of the next generation of CMB and BAO probes. The case $\Omk = -10^{-3}$ is a factor of 5 below, and still a factor of 100 above the absolute lower limit of $\Omk \sim 10^{5}$. Yet we will demonstrate that this scenario already presents very considerable challenges in terms of model confusion. Finally, the flat case allows us to investigate whether future probes can correctly determine (in a model selection sense) if the inflationary prediction is correct. In the following, we focus on the closed Universe case, because this has the added interest of a finite Universe, and therefore it allows to investigate the question of whether the Universe's spatial extent is infinite or not. In terms of parameters constraints and model selection outcomes, the conclusions are expected to hold almost unchanged for the case of fiducial Universes with $\Omk>0$, i.e. for the open case. 

\section{Results}
\label{sec:today}

\subsection{Current evidence for flatness} 

In this section we present our model comparison analysis from present--day data. Our results (obtained using a modified version of the \texttt{CosmoMC} code,~\cite{Lewis:2002ah}) are presented in Table~\ref{tab:currentdata}. 

\begin{table*}
\begin{center}\begin{tabular}{l | cccc | l |}
Data sets and models &  $\ln B_{01}$  &  $\ln B_{0-1}$ &  $p(\mdl_0 | d)$ & $p(N_U = \infty | d)$ & Notes 
\\ \hline \hline 
 & \multicolumn{5}{c|}{Astronomer's prior (flat in $\Omk$)} 
 \\ \hline 
WMAP5+BAO ($w=-1$)  		& $4.1$ &  $5.3$ & $0.98$  & $0.98$   & Moderate evidence for a flat, infinite Universe   \\ 
WMAP5+BAO+SNIa ($w=-1$)     & $4.2$ &  $5.3$ & $0.98$  & $0.98$   & Moderate evidence for a flat, infinite Universe      \\\hline
WMAP5+BAO ($w\neq -1$)       & $1.0$ &  $6.1$ & $0.74$  & $0.74$   & Weak evidence for flatness   \\
WMAP5+BAO+SNIa ($w\neq-1$) & $3.9$ &  $5.3$ & $0.98$   & $0.98$  & Moderate evidence for flatness  \\\hline
&  \multicolumn{5}{c|}{Curvature scale prior (flat in $\lomk$)} \\ \hline

WMAP5+BAO ($w=-1$) 	     & $0.4$ & $0.6$ & $0.45$  & $0.69$  & Inconclusive \\
WMAP5+BAO+SNIa ($w=-1$)     & $0.4$ & $0.6$ & $0.45$ & $0.69$   & Inconclusive \\\hline
WMAP5+BAO ($w\neq -1$)      & $-0.8$ & $0.5$ & $0.26$ & $0.42$   & Inconclusive  \\
WMAP5+BAO+SNIa ($w\neq-1$) & $0.3$ & $0.6$ & $0.44$  & $0.67$   & Inconclusive 
\\\hline			  
\end{tabular} \caption{Outcome of a three--way Bayesian model selection for the curvature of the Universe from current data and two choices of priors. For a prior choice motivated by astronomical considerations (`Astronomer's prior'), the posterior probability for a flat, infinite inflationary model ($p(\mdl_0 | d)$ column) increases from the initial $33\%$ to about $98\%$ for the most constraining data combination, even if the assumption of a cosmological constant is dropped. On the contrary, the `Curvature scale prior' returns an inconclusive model comparison, because in this case the Occam's razor effect is much reduced. The column  $p(N_U = \infty | d)$ gives the probability of the Universe being infinite. \label{tab:currentdata}}
\end{center}
\end{table*}

Starting with the Astronomer's prior case, we find moderate evidence for a flat Universe when compared with a closed model ($\ln B_{01} \approx 4$ for all cases but the WMAP5+BAO data combination with $w\neq -1$, which is discussed below). This corresponds to posterior odds of about 54:1. The evidence in favour of a flat model is stronger when it is compared with the open case, as a consequence of the fact that the posterior for $\Omk$ is slightly skewed towards values $\Omk < 0$, giving odds of order 200:1 in favour of the flat vs the open model. When compared against each other, the closed model is preferred over the open model with odds of about 4:1. Although the odds in favour of a flat Universe vs a closed one are of the same order as found in previous works (e.g.,~\cite{Kunz:2006mc} found odds of 48:1 from WMAP 3--yrs data and other constraints), one has to bear in mind that we are performing a three--way model comparison, while previous analyses have compared the flat model with arbitrarily curved ones (both open and closed). If we use the same priors as~\cite{Kunz:2006mc}, we find a Bayes factor between the flat and curved models $\ln B = 4.4$ ($\ln B = 4.6$), for $w = -1$ ($\weff \neq -1$). This translates in odds of approximately 90:1 in favour of flatness when compared with a generic curved model. Thus the latest data have improved the model comparison outcome roughly by a factor of 2. The posterior probability for an inflationary, infinite Universe is about 98\%, up from the initial 33\% from our prior choice. The above results hold true even if one relaxes the assumption of a cosmological constant for the most constraining data combination, namely the one including SNIa. However, the evidence is favour of flatness weakens considerably if one only employs WMAP5, SNIa and BAO while at the same time allowing for a non--constant dark energy equation of state ($\ln B_{01} = 1.0$). This is because the inclusion of BAO data skews the posterior for $\Omk$ to considerably negative values, thus preferring a closed Universe. Notice that for this prior the probability of a flat Universe ($p(\mdl_0 | d)$) and of an infinite Universe ( $p(N_U = \infty | d)$ ) essentially coincide, for the Occam's razor effect acts strongly against open models, as we have seen, and therefore most of the models' posterior probability is concentrated in the flat Universe.

If instead we consider the case of the Curvature scale prior (flat in $\lomk$), then the Occam's razor effect penalizing non--flat models is much weaker. This comes about because the posterior becomes flat for $\lomk \lsim -2$ and stays flat all the way to $\lomk = -5$, since for such small values of the curvature parameter, present--day data do not provide any constraint. Therefore this prior choice can be seen as more conservative in that it presents a reduced Occam's razor penalty for non--flat models. From the results in Table~\ref{tab:currentdata}, we see that for this prior choice the preference for flatness is much reduced, although $\ln B_{01}$ remains mostly positive, thus signaling a preference for the flat case. For example, the odds in favour of flatness vs closed (open) models are reduced to the order 3:2 (9:5), barring the case of $w\neq-1$ and  WMAP5, SNIa and BAO. However, these values are now below even the `weak evidence' threshold, and therefore the model comparison is inconclusive with this prior. Indeed, the posterior probability for the inflationary model (i.e., $\mdl_0$) is now only about 45\% (up from 33\% from the prior), while the probability of us living in an infinite Universe remains almost unchanged at 69\% (from about 67\% in the prior). This happens because in the light of the data, the models' probability is redistributed in such a way that the sum of the flat and open models' probability remains almost constant, despite the fact that the flat model's probability has risen and the open model's probability has been reduced (down to about 24\% from the initial 33\%).

\subsection{Constraints on the number of Hubble spheres}

\begin{figure}
\begin{center}
\includegraphics[width=\linewidth]{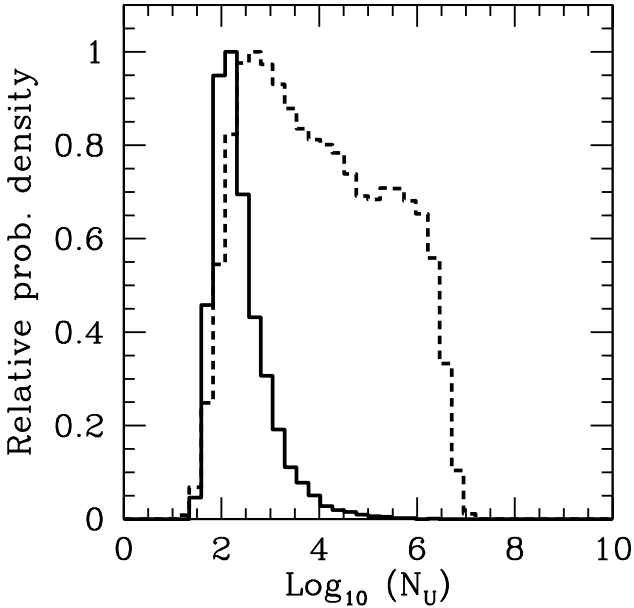}
\caption{Posterior probability distribution (normalized to the peak) for the number of Hubble spheres contained in a spatial slice (for a closed Universe) from present--day CMB+BAO+SNIa data, for the more conservative case $\weff \neq-1$ and assuming the Astronomer's prior (solid) or the Curvature scale prior (dashed).}
\label{fig:NU}
\end{center}
\end{figure}

For values $N_U < \infty$ (i.e., for closed models), the posterior probability distribution $p(N_U | d,  \mdl_1)$ is shown in Fig.~\ref{fig:NU} for both choices of priors as a function of $\log_{10}(N_U)$.  Within the class of closed models, we read off Fig.~\ref{fig:NU} that the number of Hubble spheres is constrained to lie below $N_U \sim 10^6$ for the Astronomer's prior and less than $N_U \sim 10^7$ for the Curvature scale prior. The sharp drop in the probability density for large values of $N_U$ is a reflection of the lower cut--off value chosen for the priors, $|\Omk| > 10^{-5}$, while the difference in the upper limit is a consequence of the different volume of parameter space enclosed by the two priors. The exact value of the 99\% lower limit slightly depends on the prior, as different priors allocate a different probability mass to low curvature, i.e. to large $N_U$. For the Astronomer's prior we find a 99\% lower limit (1--tail)  $N_U \gsim 4.8$, while for the Curvature scale prior this slightly increases to $N_U \gsim 6.2$ (both figures for the more conservative case where $\weff = -1$. So we conclude that, at the 99\% level, the value $N_U \gsim 5$ can be taken to be a robust lower limit to the number of Hubble spheres in the Universe. This is in good agreement with the results of the simpler analysis presented in~\cite{Scott:2006kga}, which estimated $N_U \gsim 10$.

One could also report model--averaged constraints on $N_U$, by taking into account the spread of posterior probability between the three models: 
\be
p(N_U | d) =  p(\mdl_1 | d)p(N_U | d,  \mdl_1), 
\ee
where the closed model probability, $p(\mdl_1 | d)$, can be computed from the Bayes factors reported in Table~\ref{tab:currentdata}, using the relationship
\be
p(\mdl_1 | d) = \frac{1}{1+B_{01} + B_{0-1}^{-1}}.
\ee
For the Astronomer's prior we obtain $p(\mdl_1 | d) = 0.02$ while for the Curvature scale prior $p(\mdl_1 | d) = 0.35$ for the most constraining data combination (and allowing for $w\neq -1$). Because the bulk of the model probability lies with the models where the Universe is infinite, we expect that model--averaged lower limits on $N_U$ would be {\em more} stringent than the robust limit we reported above, but also more strongly prior dependent. (A similar effect is observed for model--averaged constraints on the dark energy equation of state by~\cite{Liddle:2006kn}). For this reason, we prefer not to report model--averaged limits in this case.   

\section{Future prospects}
\label{sec:tomorrow}

We now turn to the investigation of the accuracy that future CMB and
BAO probes will achieve on $\Omk$, both from the point of view of
parameter constraints and, crucially, from the model selection perspective. Many studies have recently evaluated observational prospects using a variety of probes \citep{Knox:2006ux,Knox:2005hx,Mao:2008ug}. Here we improve on past works by analysing the results from a Bayesian model comparison viewpoint.

We assume 3 different fiducial values for $\Omk$: a flat Universe
($\Omkfid = 0$) and two possibilities for a closed Universe, namely $\Omkfid
= -10^{-3}$ (about one order of magnitude below current constraints)
and a more optimistic $\Omkfid = -5 \times 10^{-3}$. We then simulate
future CMB and BAO observations as described above. An important point is that we simulate data around the true value of the fiducial model's parameters. This is consistent with what one would expect to obtain from the average of many data realizations, and analogous to what is usually assumed with Fisher Matrix forecasts. However, from the point of view of performance prediction and model comparison it is important to stress that this choice is optimistic, in that it ignores the extra uncertainty due to the realization noise of the specific data realization that one happens to observe. 

\subsection{How flat can you get?}

\begin{figure}
\begin{center}
\includegraphics[width=\linewidth]{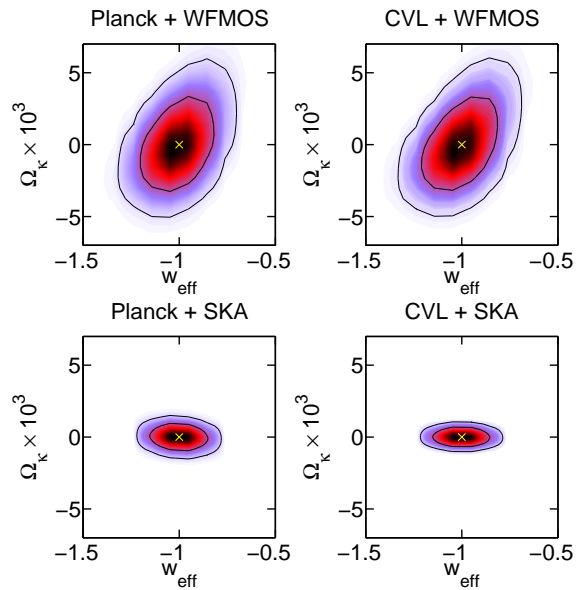}
\caption{Future constraints on curvature and the dark energy effective equation of state from various combinations of future probes, for the Astronomer's prior. Contours delimit 68\% and 95\% joint credible regions, the cross gives the fiducial value.}
\label{fig:futureconstraints}
\end{center}
\end{figure}

We first focus on the flat fiducial model, in which case no deviation from flatness should be observed (with the important caveat of realization uncertainty given above) and future probes will further tighten constraints around $\Omk = 0$. In Table~\ref{tab:flat_fiducial} we report the projected posterior $1\sigma$ constraint on $\Omk$ as well as the 99\% ($2.58\sigma$) 1--tail lower limit on $\Omk$. This quantity would be the appropriate figure to report in the case that no deviation from flatness is found and one wanted to constrain  positively closed models at the 99\% level. This limit can also be translated into the corresponding 99\% lower bound on the number of Hubble spheres, $N_U$, which is also given in Table~\ref{tab:flat_fiducial}. Combination of future CMB data with WFMOS BAO determinations will constrain curvature at the $\sim 10^{-3}$ level, with the degradation in accuracy coming from dropping the assumption of a cosmological constant being about a factor of 2. Interestingly, once Planck data are available there is not much to be gained in terms of curvature constraints from a CVL CMB experiment. An SKA--like BAO experiment will further tighten constraints by a factor of about 5, and reduce the dependency of the marginal curvature accuracy on the assumptions about the dark energy equation of state. Constraints in the $(\Omk, \weff)$ plane for the flat prior case are depicted in Fig.~\ref{fig:futureconstraints}, showing how SKA will essentially eliminate the correlation between the two parameters, leading to independent constraints on the curvature and the effective dark energy equation of state. 

This result could potentially be weakened if one allowed for a more general dark energy time dependence than we have considered here. However, \cite{Knox:2006ux} showed that WFMOS BAO constraints on $\Omk$ are remarkably robust even if one allows for an evolving dark energy of the form $w(z) = w_0 + w_a \tfrac{z}{1+z}$. This is mainly due to the extra constraining power coming from the high--redshift bin, which in our analysis played a subdominant role since we assumed that the effective equation of state is constant with redshift. Even for the more general $(w_0, w_a)$ parameterization, \cite{Knox:2006ux} found that WFMOS--like constraints on $\Omk$ are only degraded by about 50\% wrt our result (see below for further comments about the impact on model confusion). We note that our forecast for SKA--like BAO data is of the same order of the accuracy that could be achieved by a combination of weak lensing and BAO observations by the Large Synoptic Survey Telescope (LSST) when marginalizing over a more general $(w_0, w_a)$~\citep{Zhan:2006gi}. However, if one models the dark energy equation of state as a continuous function, then constraints on curvature are very considerably degraded. Even a combination of weak lensing and BAO observations by the LSST will only  achieve a relatively modest accuracy $\sim 0.017$ on $\Omk$~\citep{Zhan:2008jh}.

In terms of constraining the number of Hubble spheres, WFMOS data would increase the current lower limit $N_U \gsim 5$ by almost 2 orders of magnitude to $N_U \gsim 300-400$, while SKA--like BAO observations would further improve this by 1 order of magnitude to $N_U \gsim 2000-3000$ (all figures are given before model--averaging).

\begin{table}
\begin{center}\begin{tabular}{l | l l rl}
Probe 	   & $1 \sigma$ error on $\Omk$ &  \multicolumn{2}{c}{99\% 1--tail lower limit} \\
	  	 &   								& $\Omk$ & $N_U$   \\ \hline \hline
& \multicolumn{3}{c}{$w=-1$}\\\hline		 
Planck + WFMOS &   $1.76\cdot10^{-3}$& $-4.17\cdot10^{-3}$   &   392\\
CVL + WFMOS    &    $1.60\cdot10^{-3}$& $-3.85\cdot10^{-3}$   & 443\\
Planck + SKA      &    $5.64\cdot10^{-4}$&  $-1.34\cdot10^{-3}$   & 1970\\
CVL  + SKA          &    $4.58\cdot10^{-4}$& $-1.07\cdot10^{-3}$   &  2732
\\\hline 
& \multicolumn{3}{c}{$w \neq -1$}\\\hline	
Planck + WFMOS &  $2.22\cdot10^{-3}$  &$-4.58\cdot10^{-3}$    & 284 \\
CVL + WFMOS    &   $2.08\cdot10^{-3}$  & $-4.40\cdot10^{-3}$   & 293\\
Planck + SKA      &    $6.38\cdot10^{-4}$  &$-1.50\cdot10^{-3}$   & 1676\\
CVL  + SKA          &   $4.58\cdot10^{-4}$  & $-1.05\cdot10^{-3}$  &2723 \\\hline
\end{tabular} 
\caption{Posterior constraints on the curvature parameter $\Omk$ from future CMB and BAO probes, taking a fiducial value  $\Omkfid = 0$ and for a flat prior on $\Omk$. \label{tab:flat_fiducial}}
\end{center}
\end{table}

We now turn to the model selection question of whether future experiments will be able to determine unambiguously that the Universe is flat, should this be the case.
Results are shown in Table~\ref{tab:flat_fiducial_MS}, which gives value of $\ln B_{01}$, the Bayes factor between the (correct) flat model and the closed model (recall that $\ln B_{01}>0$ favours the flat model). The values of $\ln B_{0-1}$ are within a few percent from the ones given in the table, and hence are not displayed (the small difference comes from the fact that the dependency of the observables is not precisely symmetric in $\Omk$ around $\Omk=0$). Our findings show that all of the experiments will be able to return strong evidence ($\ln B_{01} > 5$) for the case of a flat prior on $\Omk$. This results holds true even if we relax the assumption that dark energy is in the form of a cosmological constant. 

\begin{table}
\begin{center}\begin{tabular}{l | ll | ll }
$(\Omkfid=0.0)$ &   \multicolumn{2}{c}{Astronomer's prior}  &  \multicolumn{2}{c}{Curvature scale prior}  \\
Probe			   & $w=-1$ &  $w\neq-1 $  & $w=-1$ &  $w\neq-1 $     \\ \hline \hline
%%%%% 
Planck + WFMOS & $6.0$ &  $5.9$ & $ 0.7$ & $ 0.7$\\ 
CVL + WFMOS     & $6.2$ &  $5.9$ & $ 0.8$ & $ 0.7$\\  
Planck + SKA        & $7.1$ &  $6.2$ & $ 1.0$ & $ 1.0$\\ 
CVL + SKA            & $7.5$ &  $6.3$ & $1.1$  & $1.1$ 
\\\hline			  
\end{tabular} \caption{Outcome of Bayesian model selection from future data, generated from a flat Universe. The table gives values of $\ln B_{01}$, the Bayes factor between a flat and a closed model, using a flat prior on $\Omk$ (`Astronomer's prior') or a flat prior on $\lomk$ (`Curvature scale prior'). The analysis with a flat prior on $\Omk$ gives strong evidence in favour of the flat model even when the assumption of a cosmological constant is relaxed ($w\neq -1$ column), while using a flat prior on $\lomk$ the strength of evidence is just above the `weak' threshold even for the most powerful probe (CVL+SKA). \label{tab:flat_fiducial_MS}}
\end{center}
\end{table}

\begin{figure}
\centering
\includegraphics[width=\linewidth]{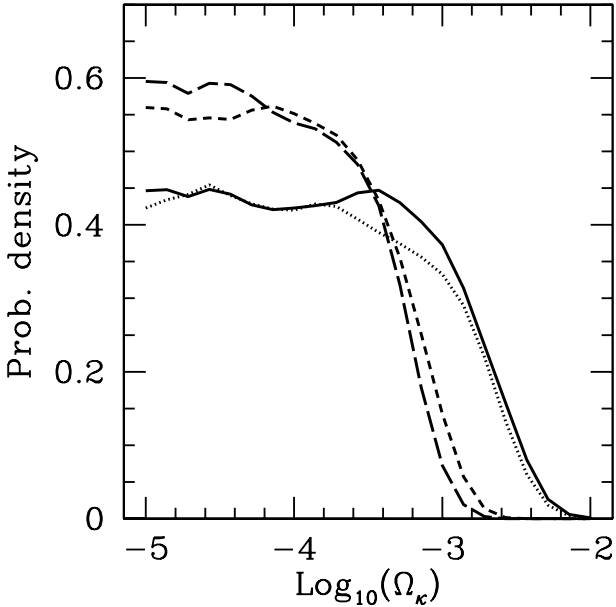}
\caption{Normalized posterior on $\lomk$ assuming a fiducial value $\Omkfid = 0$, reconstructed using a flat prior on $\lomk$ and assuming $w=-1$, for different combinations of future data. From right to left: Planck+WFMOS (solid), CVL+WFMOS (dotted), Planck+SKA (short--dashed), CVL+SKA (long--dashed). \label{fig:pdf_logp}}
\end{figure}

However, the strength of evidence is much reduced if instead one employs a prior that is flat on $\lomk$, as shown in the right--hand--side of Table~\ref{tab:flat_fiducial_MS} (`Curvature scale prior'). Even the most constraining experiments (CVL+SKA) will struggle to gather weak evidence ($\ln B_{01} > 1.0$) in favour of flatness. This comes about for two reasons. First, evidence accumulates only proportionally to the inverse error on the parameter of interest, hence in the Bayesian framework it is much easier to disprove a model (where the evidence goes exponentially in the number of sigma discrepancy with the prediction) than to confirm it. Secondly, the Occam's razor effect penalising non--flat models is much reduced under the assumption of a flat prior in $\lomk$, as the net result of two opposite effects. Fig.~\ref{fig:pdf_logp} shows the posterior pdf on $\lomk$ for the different probes (assuming $w=-1$). It is clear that for values of $\lomk \ll -3$ the posterior becomes essentially flat, reflecting the inability of the experiment to measure a curvature value much below that threshold. At the same time, the volume of parameter space enclosed by a prior flat in log space is increased with respect to the case of a linear scale. This ought to favour the simpler (flat) model. But the posterior volume is also increased, and therefore the net effect is to reduce the overall Occam's razor penalty term (which goes as the log of the ratio between the two volumes), hence the strength of evidence in favour of flatness is reduced. 

\subsection{The danger of model confusion}
\label{sec:modconf} 

We now turn to the case where the fiducial model is closed, and evaluate the resulting evidence from future data. In this case, a successful model comparison should return a preference for the closed model. 

We start with the more optimistic case of a relatively large fiducial value for the curvature parameter,  $\Omkfid = -5\times10^{-3}$, roughly a factor of 2 below present--day constraints. We give results in Table~\ref{tab:closed_fiducial_optimistic}, which shows that the flat prior on $\lomk$ always returns the correct model comparison (negative values of $\ln B_{01}$ in the table). However, the Astronomer's prior incorrectly penalizes curved models when the constraining power of the data is insufficient to overturn the Occam's razor effect (positive values in the table). This `model confusion' effect is worse when the equation of state of dark energy is allowed to change, in which case e.g. Planck+WFMOS would incorrectly gather moderate evidence in favour of flatness. With CMB and WFMOS data, the analysis is subject to model ambiguity, i.e. the result depends on the choice of prior. In order to recover the correct model selection outcome unambiguously, one needs SKA--quality BAO data to complement the CMB distance probes (negative values of  $\ln B_{01}$ for both priors).

\begin{table}
\begin{center}\begin{tabular}{l | ll | ll }
$(\Omkfid=-5\times10^{-3})$ &   \multicolumn{2}{c}{Astronomer's prior}  &  \multicolumn{2}{c}{Curvature scale prior}\\
Probe			   & $w=-1$ &  $w\neq-1 $  & $w=-1$ &  $w\neq-1 $     \\ \hline \hline
%%%%% 
Planck + WFMOS & $1.3$     &  $2.6$         & $ -3.5$ & $-1.7$\\ 
CVL + WFMOS     & $0.4$     &  $2.0$         & $ -4.5$ & $-2.0$\\  
Planck + SKA        & $-34$    &  $-22$         & $ -50$  & $-40$\\ 
CVL + SKA            & $-55$    &  $-50$        &  $ -65$  & $-58$ \\\hline			  
\end{tabular} \caption{Outcome of Bayesian model selection from future data, generated from a closed Universe,  $\Omkfid = -5\times10^{-3}$. The table gives values of $\ln B_{01}$, the Bayes factor between a flat and a closed Universe. Negative values correctly favour the closed case, while positive values wrongly favour the flat case, giving rise to model confusion. SKA--quality BAO data are required to overcome model confusion independently of the choice of prior. \label{tab:closed_fiducial_optimistic}}
\end{center}
\end{table}

The danger of model confusion becomes stronger the smaller the fiducial value one chooses for $\Omkfid$. We illustrate this by considering our third fiducial model, namely a closed Universe with $\Omkfid = -10^{-3}$, which is about one order of magnitude below current limits but still two orders of magnitudes above the fundamental fluctuation limit. The model comparison outcome is given in Table~\ref{tab:closed_fiducial_low}, which shows that this case results in widespread model confusion for the Astronomer's prior, for which the flat Universe is incorrectly preferred with moderate to strong evidence by all combinations of probes. For the Curvature scale prior, instead, the outcome is always inconclusive, even though the CVL+SKA combination does reach the correct conclusion, albeit with evidence which falls short even of the `weak' threshold.

\begin{table}
\begin{center}\begin{tabular}{l | ll | ll }
$(\Omkfid=-10^{-3})$ &   \multicolumn{2}{c}{Astronomer's prior}  &  \multicolumn{2}{c}{Curvature scale prior}\\
Probe			   & $w=-1$ &  $w\neq-1 $  & $w=-1$ &  $w\neq-1 $     \\ \hline \hline
%%%%%%
Planck + WFMOS & $5.6$  & $5.5$ & $0.6$ & $0.6$\\ 
CVL + WFMOS     & $5.6$ &  $5.2$ & $0.4$ & $0.6$\\  
Planck + SKA        & $5.0$ &  $5.2$ & $0.0$ & $0.1$\\ 
CVL + SKA            & $4.4$ &  $4.4$ & $-0.6$ & $-0.6$ \\\hline			  
\end{tabular} 
\caption{As in Table~\ref{tab:closed_fiducial_optimistic}, but for a fiducial value $\Omkfid = -10^{-3}$. For such a small value of the curvature, only the CVL+SKA data combination achieves the correct model selection (albeit with undecided odds) and this only when employing the curvature scale prior. All other cases are subject to model confusion. \label{tab:closed_fiducial_low}}
\end{center}
\end{table}

Some comments are in order about the robustness of those results with respect to changes in the assumed dark energy model. In particular, an evolving dark energy component could mimic to an extent the effect of curvature~\citep{Clarkson:2007bc}, and this would lead to increased uncertainty in the curvature parameter and thus to increased model confusion. To estimate the impact of this effect, we have repeated the analysis for a subset of the cases discussed above, but marginalising over a 2--parameters dark energy equation of state of the form  $w(z) = w_0 + w_a \tfrac{z}{1+z}$. In this case, the values of $\ln B_{01}$ are reduced by $\sim 10\%-20\%$ with respect to the case where a $\weff \neq -1$ model was assumed. This change is not large enough to modify in a significant way the outcome of model selection reported in Tables~\ref{tab:flat_fiducial_MS}--\ref{tab:closed_fiducial_low}. Therefore we conclude that assuming a more general dark energy equation of state does not impact very strongly on our results about the danger of model confusion.

\subsection{Avoiding model confusion}

In the light of the findings in the previous section, it is interesting to estimate the required accuracy on $\Omk$ in order to ensure that future probes will not be subject to model confusion. For a given fiducial value of $|\Omk|>10^{-5}$, we wish to estimate the accuracy needed so that the model comparison correctly favours the closed model over a flat one independently of the choice of prior. 

This can be achieved by using a Gaussian approximation to the future likelihood and employing the SDDR to estimate the Bayes factor between the closed and open model that a future experiment would obtain. We start by considering the case of the Astronomer's prior.  We assume that the marginal likelihood of $\Omk$ is approximately described by a Gaussian with mean $\Omkfid$ (i.e., centered on the fiducial value\footnote{As discussed above, this neglects the realization noise and is therefore equivalent to an ensemble--averaged forecast, analogous to what is usually done for Fisher matrix forecasts. However, numerical investigations suggest that realization noise is a subdominant source of uncertainty in this context (Andrew Liddle, private communication).}) and variance $\Sigma^2$. Then adopting the Astronomer's prior, the Bayes factor between the flat and the closed model is given by, from Eq.~\eqref{eq:savagedickey}.
 \be \label{eq:B01flat}
 \ln B_{01} \approx -\ln \frac{\Sigma}{\Delta \Theta} - f_A(\Omkfid, \Sigma) - \frac{1}{2}\frac{(\Omkfid)^2}{\Sigma^2} 
 \ee
where $\Delta \Theta =1$ is the width of the Astronomer's prior on the curvature parameter and the last term of the right--hand--side is defined as  
 \be
 f_A(\Omkfid, \Sigma) \equiv \ln \sqrt{\frac{\pi}{2}} \[ \erf \left( \frac{\Delta\Theta-|\Omkfid|}{\sqrt{2}\Sigma} \right) + \erf \left( \frac{|\Omkfid|}{\sqrt{2}\Sigma} \right) \]
 \ee
and $\erf(x)$ denotes the error function, 
 \be
 \erf(x) \equiv \frac{2}{\pi}\int_0^x \exp(\tau^2)\dr \tau .
 \ee
The function $f_A$ accounts for the upper and lower limits in the Astronomer's prior distribution when computing the evidence. It is easy to see that when the posterior is sharply localized within the prior, i.e. for $\Omkfid/\Sigma \gg 1$ it follows that  $ f_A(\Omkfid, \Sigma) \rightarrow \frac{1}{2}\ln2\pi$. Thus in Eq.~\eqref{eq:B01flat}, the first two terms on the right--hand--side represent the Occam's razor effect (notice that $-\ln\Sigma/\Delta\Theta > 0$, thus favouring the flat model) , while the last term describes the relative quality of fit between the closed and flat model. We have checked the accuracy of the approximation of Eq.~\eqref{eq:B01flat} against the full numerical results in Tables~\ref{tab:closed_fiducial_optimistic} and \ref{tab:closed_fiducial_low} adopting the error estimates given in Table~\ref{tab:flat_fiducial} and we have found it to be excellent, with an accuracy of a few percent. 

\begin{figure}
\begin{center}
\includegraphics[width=\linewidth]{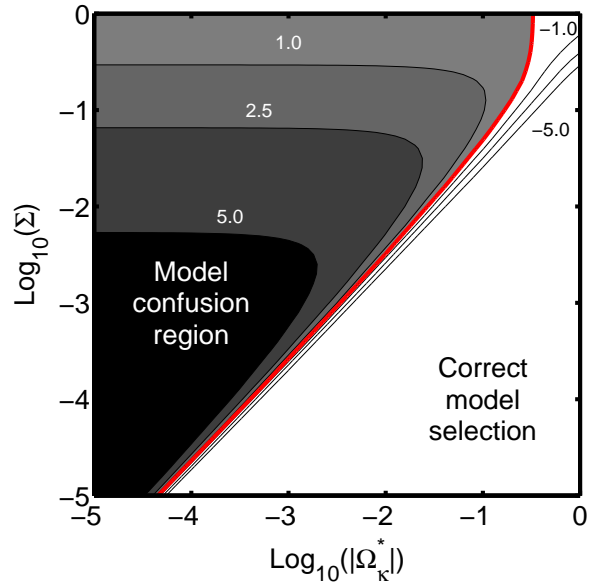}
\caption{Bayes factor from future data for the Astronomer's prior as a function of the true value of the curvature parameter $\Omkfid$ and the future marginal accuracy on $\Omk$, $\Sigma$. The red, thick line separates regions of model confusion (above the line, shaded, $\ln B_{01} > 0$, wrongly favouring a flat Universe) from regions of correct model selection ($\ln B_{01} < 0$, white, correctly returning a preference for a closed Universe). The contours denote increasing levels of evidence, with values of $\ln B_{01}$ as labelled. The contours below the red line delimit regions of weak, moderate and strong preference for the closed Universe (from top to bottom).
\label{fig:B01flat}}
\end{center}
\end{figure}

The result is plotted in Fig.~\ref{fig:B01flat}, where the red, thick line is the contour level  $\ln B_{01} = 0$ which separate the region where the model comparison correctly favours a closed Universe (bottom right corner, in white) from the `model confusion' region, where the flat Universe is incorrectly preferred due to the Occam's razor effect (shaded region above the red line). For a given value of the curvature parameter on the horizontal axis, the red line thus gives the required marginal accuracy on $\Omk$ to avoid model confusion. We will come back below to discussing what this means in terms of the required discovery threshold. The unfilled contours below the red line denote values $\ln B_{01} = -1.0, -2.5, -5.0$ (weak, moderate and strong evidence for curvature, respectively, from top to bottom in the figure). Because the evidence against the null hypothesis of a flat Universe grows exponentially in the tails of the distribution, those contours are relatively close to the $\ln B_{01} = 0$ threshold. This means that a relatively modest increase in accuracy can lead to `strong' evidence in favour of curvature. The situation is not symmetric with respect to the null hypothesis: the evidence increases only linearly with the accuracy in case of a null result, hence it takes a much larger accuracy to accumulate evidence in favour of the null. This is reflected by the larger spacing between the evidence  contours in the model confusion region.

\begin{figure}
\begin{center}
\includegraphics[width=\linewidth]{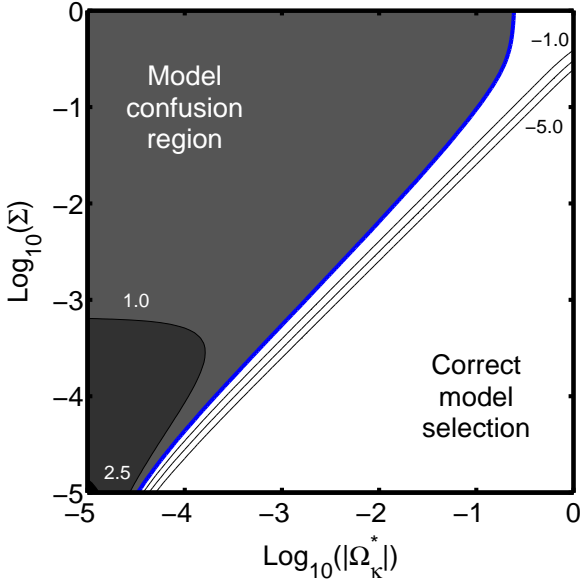}
\caption{As in Fig.~\ref{fig:B01flat}, but for the Curvature scale prior. The blue, thick line separates regions of model confusion (above the line) from regions of correct model selection ($\ln B_{01} < 0$). The shaded areas denote regions of increasing model confusion, from light to dark. \label{fig:B01log}}
\end{center}
\end{figure}

Turning now to the case of the Curvature scale prior, the Bayes factor can be computed in an analogous fashion, by replacing the flat prior on $\Omk$ by the prior equivalent to a flat prior on $\lomk$, namely $p(\Omk) = M/\Omk$ (for $-1  \leq \Omk \leq -10^{-5}$) and $M$ is a normalization constant. The Bayes factor can then be computed numerically using the SDDR, Eq.~\eqref{eq:savagedickey}. The resulting outcome for model selection is shown in Fig.~\ref{fig:B01log}, where the blue, thick line again separates region of correct model selection from regions of model confusion, as a function of the fiducial value for the curvature and of the marginal accuracy. By comparing with Fig.~\ref{fig:B01flat}, we notice that the Curvature scale prior is less subject to model confusion than the Astronomer's prior, since for the former the strength of evidence in favour of a flat Universe is lower in the model confusion region and it barely reaches the `moderate' evidence threshold. Furthermore, the blue line is always above the red line (see Fig.~\ref{fig:ModAmb}), which means that model confusion for the Curvature scale prior is avoided with less stringent requirements on the marginal accuracy $\Sigma$ than for the Astronomer's prior.   

\subsection{Limits to the knowability of the the geometry}

\begin{figure}
\begin{center}
\includegraphics[width=\linewidth]{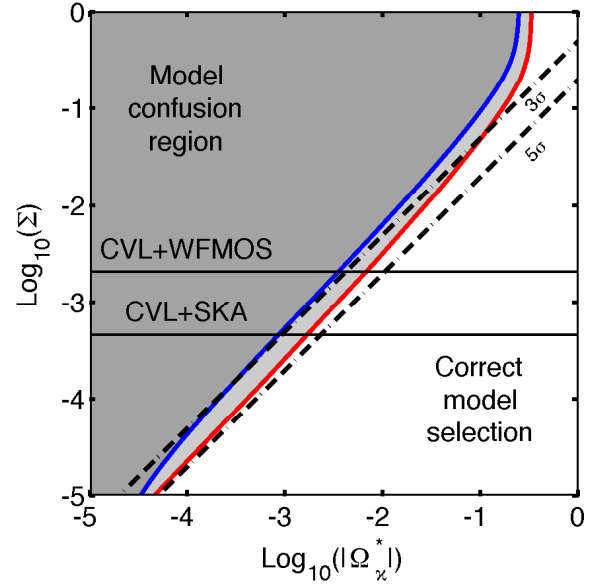}
\caption{The red (blue) line delimits regions of model confusion (above the line) for the Astronomer's prior (Curvature scale prior). The light--shaded region between the red and blue lines is a zone of `model ambiguity', where the model comparison results depends on the prior. The diagonal, dotted lines denote approximate regions of 3$\sigma$ and 5$\sigma$ (from top to bottom) discovery for a given true value of the curvature parameter, $\Omkfid$. A 5$\sigma$ discovery threshold guarantees an ambiguity-- and confusion--free model comparison outcome. The horizontal lines give the expected accuracy of future CMB and BAO probes.  \label{fig:ModAmb}}
\end{center}
\end{figure}

The comparison between the two priors is further investigated in Fig.~\ref{fig:ModAmb}, where we plot the contours separating the model confusion region for both priors (red for the Astronomer's prior and blue for the Curvature scale prior). The dark shaded region labeled `model confusion' leads to an erroneous model comparison result for both priors, while the light shaded region between the two lines is a zone of `model ambiguity' --- where the outcome of model comparison depends on the choice of prior. In such a case, better data (i.e., smaller $\Sigma$) are required in order to resolve the ambiguity. It is interesting to investigate the accuracy necessary to obtain an ambiguity-- and confusion--free model selection. In Fig.~\ref{fig:ModAmb}, the diagonal, dashed lines represents approximate 3 and 5$\sigma$ detection thresholds (from top to bottom). We can see that, except for fairly large values of $\Omkfid \gsim 0.1$, a 3$\sigma$ `detection' is subject to both model ambiguity and model confusion. On the other hand, a 5$\sigma$ detection leads to an unambiguous and correct model choice all the way down to $\Omkfid \gsim 7\times 10^{-5}$.

This is further substantiated by the results tabulated in Table~\ref{tab:error_required}, giving the required accuracy (both in absolute value and number of $\sigma$ discovery) to achieve moderate or strong evidence under both priors, for a few representative choices of the fiducial curvature value. Under the Astronomer's prior, moderate evidence in favour of curvature requires a $\sim 4\sigma$ detection, while for strong evidence a  $\sim 5\sigma$ detection is necessary. As mentioned above, the Curvature scale prior is less demanding due to its reduced Occam's razor effect: moderate evidence is achieved with a $\sim 3.2\sigma$ detection threshold, while strong evidence is obtained at $\sim 4\sigma$. This of course comes at the price of a much reduced evidence in favour of a flat Universe if that is indeed the true model, as discussed in connection with the results presented in Table~\ref{tab:flat_fiducial_MS}.

In conclusion, our results imply that a $5\sigma$ detection threshold ought to be recommended in order to obtain a secure and ambiguity--free model selection. 
It is perhaps amusing that a full Bayesian treatment of the problem concludes that the $5\sigma$ detection threshold traditionally used in particle physics (with its frequentist framework) ought to be employed. 

\begin{table*}
\begin{center}\begin{tabular}{l | ll | ll }
True value  &   \multicolumn{2}{c}{Astronomer's prior}  &  \multicolumn{2}{c}{Curvature scale prior}  \\
			   &  moderate evidence   & strong evidence   &  moderate evidence   & strong evidence    \\ \hline \hline
%%%%% 
$\Omkfid = -5 \times 10^{-3}$  & $1.23\times10^{-3}$ $(4.06\sigma)$  & $1.06\times10^{-3}$ $(4.66\sigma)$ & $1.57\times10^{-3}$ $(3.16\sigma)$ & $1.26\times10^{-3}$ $(3.95\sigma$)  \\
$\Omkfid = -10^{-3}$   &$2.23\times10^{-4}$ $(4.42\sigma)$  &$2.00\times10^{-4}$ $(4.97\sigma)$  & $3.12\times10^{-4}$ $(3.17\sigma)$ &  $2.50\times10^{-4}$ $(3.95\sigma)$\\
$\Omkfid = -10^{-4}$   &$2.00\times10^{-5}$ $(4.48\sigma)$  &$1.82\times10^{-5}$ $(4.93\sigma)$  & $2.79\times10^{-5}$ $(3.22\sigma)$ &  $2.26\times10^{-5}$ $(3.99\sigma)$
\\\hline			  
\end{tabular} \caption{Required accuracy on the marginal error on $\Omk$ to avoid model confusion and to achieve different thresholds of evidence in favour of a closed Universe for the two priors considered in the text (absolute value and relative number of $\sigma$ in parenthesis). A 5$\sigma$ discovery threshold guarantees an ambiguity-- and confusion--free model comparison outcome down to $|\Omkfid| = 10^{-4}$. \label{tab:error_required}}
\end{center}
\end{table*}

Finally, we can revise the conclusion about the fundamental limit to the knowability of the geometry of the Universe. It is usually argued that this is of order $|\Omk|\sim 10^{-5}$, because this is the typical size of curvature fluctuations due to primordial inhomogeneities. However, Fig.~\ref{fig:ModAmb} shows that model confusion sets in for value of the curvature $|\Omkfid| \lsim 10^{-4}$, which means that if the true value of the curvature is below this threshold we will not be able to gather evidence for it. We conclude that the fundamental limit to our ability to detect the curvature of the Universe (if present) is of the order $|\Omk|\sim 10^{-4}$, which is an order of magnitude greater than previous estimates. Below that value, the Occam's razor arguments inbuilt into Bayesian model selection imply that we ought to revert to preferring a flat Universe. Therefore if the curvature is in the `undecidable interval' $10^{-5} \leq |\Omk| \lsim 10^{-4}$ no amount of data will be able to determine that the Universe is non--flat.

\section{Conclusions}
\label{sec:conclusions}

We have subjected the geometry of the Universe to a detailed scrutiny from a model comparison perspective, performing a three--way model selection with two physically motivated priors. We found that present--day data provide up to moderate evidence in favour of flatness (maximum odds of 66:1) for a specific choice of prior (the Astronomer's prior) and assuming that dark energy is a cosmological constant. A Curvature scale prior and a relaxation of the assumption on the nature of dark energy weaken this result considerably, giving only inconclusive odds of less than 3:2  in favour of flatness. Correspondingly, the probability that the Universe is infinite lies in the range from 67\% to 98\%, depending on assumptions. If the Universe is not infinite, we have found a robust lower limit to the number of Hubble spheres, $N_U \gsim 5$. 

We have discussed the prospects for future CMB and BAO probes to determine with strong evidence the geometry of the Universe. CMB data coupled with WFMOS BAO observations will achieve an accuracy on $\Omk$ of the order $\sim 1-2\times 10^{-3}$, while SKA--like BAO data will further increase the accuracy to $\sim 4-6\times 10^{-4}$. Allowing for the effective equation of state of dark energy to be different from $-1$ (although constant in redshift) will not significantly decrease the accuracy with which CMB+SKA data will determine $\Omk$. 

Finally, we have shown that a model selection perspective places much more taxing requirements on the accuracy of future datasets than one would naively assume. In particular, a 5$\sigma$ detection threshold is recommended in order to avoid both model confusion and model ambiguity in the determination of the geometry. However, if the value of the curvature parameter is smaller than $\sim 10^{-4}$ we found that no amount of observations will be able to decide on the true geometry of the Universe. Achieving this lower limit will require an improvement of another factor of 20 over what a CVL CMB experiment with an SKA--like BAO probe will obtain. This might be feasible once other, orthogonal datasets such as weak lensing and SNIa observations are added to the likelihood, although it will be a formidable challenge to control systematics at this level of statistical accuracy.

\bigskip

\textit{Acknowledgements} The authors would like to thank Andrew Jaffe for many useful discussions. 
M.V. is supported by the Raffy Manoukian Scholarship and partially supported by the Philip Wetton Scholarship at Christ Church, Oxford. 
R.T. was partially supported by the Royal
Astronomical Society, by STFC and by St Anne's College, Oxford.
 We acknowledge the use of the Legacy Archive
for Microwave Background Data Analysis (LAMBDA). Support for
LAMBDA is provided by the NASA Office of Space Science.

%\bibliographystyle{mn2e}
%\bibliography{../BasicBiblio,../AdditionalRefs}
\newcommand{\bibfont}{\fontsize{10}{12}\selectfont} \newcommand{\noopsort}[1]{}
  \newcommand{\printfirst}[2]{#1} \newcommand{\singleletter}[1]{#1}
  \newcommand{\switchargs}[2]{#2#1}

\end{document}